\documentclass[reprint,
 amsmath,amssymb,
 pra,
]{revtex4-1}

\pdfoutput=1

\usepackage{graphicx}
\usepackage{dcolumn}
\usepackage{bm}
\usepackage{soul}

\usepackage{esdiff}
\usepackage{esvect}
\usepackage{physics}
\usepackage{xcolor}

\newcommand{\eq}[1]{
\begin{equation}\begin{aligned}
#1
\end{aligned}\end{equation}}

\DeclareMathOperator{\chitwo}{ {\chi}^{(2)} }
\DeclareMathOperator{\chithree}{ {\chi}^{(3)} }

\DeclareMathOperator{\eps}{\varepsilon}

\begin{document}

\preprint{APS/123-QED}

\title{Raman solitons in waveguides with simultaneous quadratic and Kerr nonlinearities}

\author{William R. Rowe}
\author{Dmitry V. Skryabin}
\author{Andrey V. Gorbach}%
 \email{A.Gorbach@bath.ac.uk}
\affiliation{ Centre for Photonics and Photonic Materials, Department of Physics, 
University of Bath, Bath BA2 7AY, United Kingdom
}%

\date{\today}

\begin{abstract}
We analyse Raman-induced self-frequency shift in two-component solitons supported by both quadratic and cubic nonlinearities. Treating Raman terms as a perturbation, we derive expressions for soliton velocity and frequency shifts of the fundamental frequency and second harmonic soliton components. We find these predictions compare well with simulations of soliton propagation. We also show that Raman shift can cause two-component solitons to approach the boundary of their own existence and subsequently trigger soliton instabilities. In some cases these instabilities are accompanied by an almost complete transfer of power to the second harmonic, and emergence of a single-component Kerr solitonic pulse.
\end{abstract}

\maketitle


\section{Introduction}
Intrapulse Raman scattering (IRS) arises in ultrashort pulses with sufficient spectral breadth, allowing stimulated Raman scattering between frequencies within the pulse \cite{Agrawal2013NonlinearOptics}.  IRS in the regime soliton of propagation in materials with Kerr nonlinearity is known to lead to the continuous downshift of the central soliton frequency. This phenomenon is commonly referred to as the soliton self-frequency shift (SFS) \cite{Agrawal2013NonlinearOptics, Kivshar2003ChapterSolitonsb, Dudley2010SupercontinuumFibers}. In systems with cubic ($\chithree$) nonlinearity SFS can be described using a perturbation expansion as an adiabatic transformation of soliton frequency and velocity \cite{Gordon1986TheoryShift, Kodama1987NonlinearGuide}. This description has been crucial in the explanation of supercontinuum generation (SCG) in optical fibres \cite{Skryabin2003SolitonFibers,Biancalana2004TheoryFibers,Skryabin2010TheoryWaves} and has been shown to work well even in cases of strong Raman response, for example in air-core fibres \cite{Gorbach2008SolitonFibers}.

Nonlinear pulse dynamics in systems with quadratic ($\chitwo$) and $\chithree$ nonlinearity have become the subject or research recently, in particular prompted by the emergence of LN on insulator waveguides \cite{Guo2015SupercontinuumMatching, Phillips2011SupercontinuumWaveguides, Guo2013NonlinearMedia}.
These waveguides are an exciting new platform for nonlinear optics \cite{Poberaj2012LithiumDevices, Boes2018StatusCircuits}. They combine the broadband transparency and high quadratic ($\chitwo$) nonlinearity of LN with the dispersion engineering of high index-contrast nano-waveguides \cite{Poberaj2012LithiumDevices, Boes2018StatusCircuits}. Together these aspects allow for generation, at experimentally achievable powers, of two-component temporal solitons \cite{Rowe2019TemporalNanowaveguides}. Two-component solitons, sometimes called two-colour solitons, are solitonic excitation's consisting of two pulses, one at a fundamental frequency (FF) and one at its second harmonic (SH), locked together as they propagate \cite{Buryak2002}. This requirement that both components propagate together with same group velocity means that two-component solitons are inherently more fragile than their Kerr counterparts \cite{Buryak2002, Rowe2019TemporalNanowaveguides}. However they can be excited in normal dispersion, meaning two-component solitons may exist across the spectral range that does not allow  bright Kerr solitons \cite{Buryak2002, Rowe2019TemporalNanowaveguides}. Two-component solitons have been studied extensively in both $\chitwo$ only and $\chitwo + \chithree$ systems \cite{Schiek1993NonlinearStructures, Buryak2002, Buryak1995OpticalNonlinearities}. 

LN is known to have a strong Raman response \cite{Bache2012ReviewResponse} and recent studies into LN nano-waveguides suggest the existence of two-component Raman solitons \cite{Guo2015SupercontinuumMatching, Phillips2011SupercontinuumWaveguides, Guo2013NonlinearMedia}. So far however, no theory has been developed for IRS in such systems. It is therefore the purpose of this work to analyse two-component Raman solitons, using LN nano-waveguides as an illustrative example. In particular we would like to explore whether the delicate balance between the two components with their different dispersion's can be preserved when accounting for the effects of Raman. 

\section{Model}

We consider a waveguide with a fixed cross-section in the $x$-$y$ plane perpendicular to the direction of propagation $z$.  The FF and SH fields are expressed as $A_f\exp(i\beta_f z-i\omega_f t)$ and $A_s\exp(i\beta_s z-i\omega_s t)$, respectively.
Here,  $\omega_s=2\omega_f$, $A_f$ and $A_s$ are the field envelopes,  $t$ is time and $z$ is distance along the waveguide. $|A_{f,s}|^2$ are scaled to be measured in Watts. $\beta_f$ and $\beta_s$ are the propagation constants defined as,
\eq{
\beta_k (\omega) = \sum_{m=0} \frac{[\omega - \omega_k]^m}{m!} \beta_{km}, \qquad
\beta_{km}=\diffp[m]{\beta_k}{\omega}\bigg|_{\omega_k},
}
where $k$ is either $f$ or $s$ denoting the FF or SH respectively.  

We normalise these fields by defining the dispersion length of the FF component, $z_d = 2t_0^2/|\beta_{f2}|$ where $t_0$ is a characteristic time scale, typically the FF pulse width. Our amplitudes therefore become the normalised envelope functions, $U_f = \sqrt{2} \rho_2 z_d A_f$ and $U_s = \rho_2 z_d A_s$. $\rho_2$ is the effective $\chitwo$ nonlinearity coupling the two modes, which we define later. Propagating along the waveguide these field envelopes evolve as,
\eq{\label{e:model}
i\partial_\xi U_f +& r_2 \partial_\tau^2 U_f + U_f^* U_s e^{i\kappa \xi} \\+& U_f\big[\alpha_f|U_f|^2 + \alpha_c|U_s|^2\big] = \eps_f(U_f,U_s),\\
i\partial_\xi U_s +& is_1\partial_\tau U_s + s_2 \partial_\tau^2 U_s + \frac{U_f^2}{2} e^{-i\kappa \xi} \\ +& U_s\big[\alpha_s|U_s|^2 + \alpha_c|U_f|^2\big] = \eps_s(U_f,U_s).
}
where we have defined the normalised coordinates $\xi = z/z_d$, in the direction of propagation and $\tau = [t - \beta_{f1} z]/t_0$, the transverse time coordinate moving with the group velocity of the fundamental field. 

We define the normalised dispersion parameters,
\eq{
r_2 = -\frac{z_d}{2 t_0^2}\beta_{f2}, \qquad
s_2 = -\frac{z_d}{2 t_0^2}\beta_{s2},
}
 the walk-off parameter, $s_1 = [\beta_{s1} - \beta_{f1}] z_d/t_0$ and the phase-mismatch parameter, $\kappa = [\beta_s(\omega_s) - 2 \beta_f(\omega_f)] z_d$.
The coefficients $\alpha_k$ are defined by the balance between the effective $\chithree$ and $\chitwo$ nonlinearities and dispersion length $z_d$ with,
\eq{
\alpha_f = \frac{3\rho_{3,f}}{2\rho_2^2 z_d}, \,\,\,\,
\alpha_s = \frac{3\rho_{3,s}}{\rho_2^2 z_d},  \,\,\,\,
\alpha_c = \frac{3\rho_{3,c}}{\rho_2^2 z_d}.  \,\,\,\,
}
$\rho_{3,f}$, $\rho_{3,s}$, $\rho_{3,c}$ are the effective $\chithree$ nonlinearities in the FF mode, SH mode and between the two modes respectively. We note that $\alpha_{f,s,c}$ are set both by the waveguide geometry via $\rho_2$ and $\rho_{3,f,s,c}$ and the input pulse parameters through $t_0$ via $z_d$. Effective nonlinearities are calculated taking overlap integrals of the appropriate spacial mode profiles with the nonlinear material \cite{Cai2018HighlyNano-waveguides, Gorbach2016PerturbationGeneration},
\eq{
\rho_2 = \frac{ \eps_0 \omega_f}{4 N_f\sqrt{N_{s}}}\int \vv{\bm{e}}_{s} \cdot  \hat{\chi}^{(2)} \vdots \vv{\bm{e}}_f^2  \mathrm{d}\Omega,\\
\gamma_{kjpl} = \frac{\epsilon_0 \omega_k}{16\sqrt{N_kN_jN_pN_l} } \int \vv{\bm{e}}^*_k \cdot \hat{\chi}^{(3)} \vdots \vv{\bm{e}}_j^* \vv{\bm{e}}_p \vv{\bm{e}}_l   \dd \Omega,
}
where electric field profiles of FF and SH modes are ${\vv{\bm{e}}}_{f}$ and ${\vv{\bm{e}}}_{s}$ respectively. $N_k$ is the normalisation factor for the mode $k$. $\hat{\chi}^{(2)}$ and $\hat{\chi}^{(3)}$ are the material $\chitwo$ and $\chithree$ nonlinear tensors respectively and $\Omega$ is the cross-sectional area of the nonlinear material in the waveguide. $\rho_{3,f} = \gamma_{ffff}$, $\rho_{3,s} = \gamma_{ssss}$ and $\rho_{3,c} = 2\gamma_{fssf} = \gamma_{sffs}$. 

\begin{figure}[b]
    \centering
    \includegraphics[]{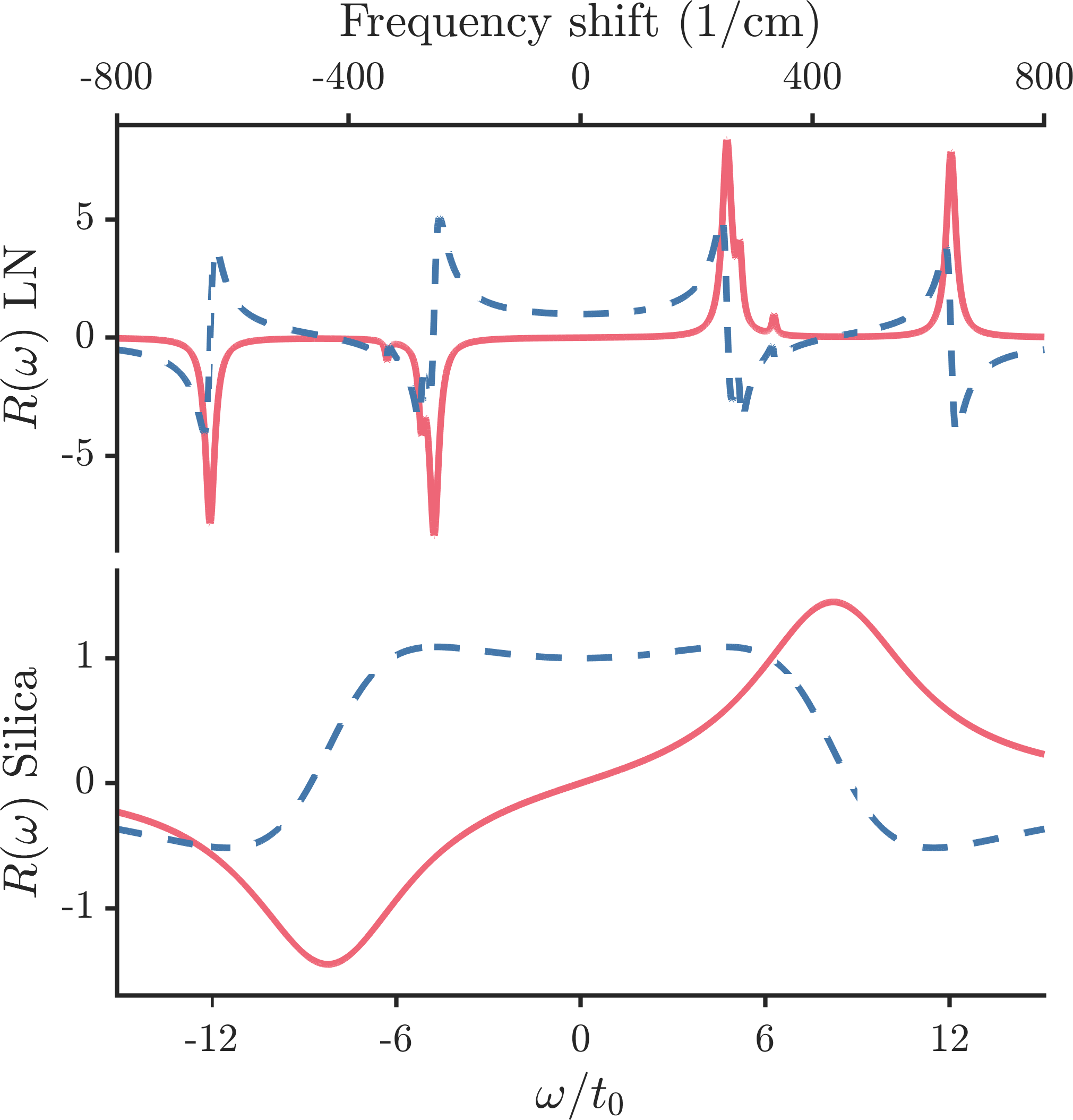}
    \caption{ Real (dashed blue) and imaginary (solid red) parts of the Raman response spectra, $R(\omega)$, for LN and silica in the upper and lower plots respectively.  }
    \label{fig:Raman_response}
\end{figure}

On the right hand side of Eq. \eqref{e:model} we have additional terms, $\eps_f$ and $\eps_s$, which must be zero for perfect soliton solutions to exist in the system (see seciton \ref{s:solitons}). We use these terms to include the delayed Raman response into our model by setting,
\eq{
\eps_f(U_f,U_s) = &  f_R U_f  \int_{-\infty}^{\infty} [ \delta(\tau' -\tau) - R(\tau' -\tau)  ]\\
&\,\,\,\,\,\,\,\,\,\,\,\,\,\,\,\,\,\,\,\, [\alpha_f |U_f(\tau')|^2 + \alpha_c |U_s(\tau')|^2] \dd \tau' ,\\
\eps_s(U_f,U_s) = &  f_R U_s  \int_{-\infty}^{\infty} [ \delta(\tau' -\tau) - R(\tau'-\tau)] \\
&\,\,\,\,\,\,\,\,\,\,\,\,\,\,\,\,\,\,\,\, [\alpha_s |U_s(\tau')|^2 + \alpha_c |U_f(\tau')|^2] \dd \tau' ,
}
where $\delta(\tau)$ is the Dirac delta function and $f_R$ is the fraction of the total material $\chithree$ response that is attributed to delayed Raman response which for LN we use $f_R =  0.5$ \cite{Bache2012ReviewResponse, Guo2013NonlinearMedia}. For silica the Raman fraction is known to be $f_R=0.18$ \cite{Skryabin2010TheoryWaves}. The Raman response function of the nonlinear material,
\eq{
R(\tau) =& \Theta(\tau) \sum_{n=1}^N f_n  \frac{\tau_{1,n}^2 + \tau_{2,n}^2}{\tau_{1,n}\tau_{2,n}^2} \sin( \frac{\tau}{\tau_{1,n}}) e^{-\tau/\tau_{2,n}},
}
where $\Theta(\tau)$ is the Heaviside step function \cite{Gorbach2008SolitonFibers}.
$f_n$ is the fractional contribution of the $n^{\rm{th}}$ resonance to the overall Raman response function. The sum is performed over the $N$ resonances of the material each of which has its own characteristic response times $\tau_1$ and $\tau_2$.
For LN we model $R(\tau)$ using $N=4$ resonances with $\tau_1 = [21, 19.3, 15.9, 8.3]\rm{fs}/t_0$, $\tau_2 = [544, 1021, 1361, 544]\rm{fs}/t_0$ and $f_n = [0.635, 0.105, 0.02, 0.24]$ \cite{Bache2012ReviewResponse}. In Silica $R(\tau)$ is typically modeled with $N=1$ resonances with $\tau_1 = 12.2\rm{fs}/t_0$, $\tau_2 = 32\rm{fs}/t_0$ and $f_n = 1$ \cite{Gorbach2007TheoryFibers}. The resulting Raman response of both LN and silica are plotted in figure \ref{fig:Raman_response} in the frequency domain.

\section{two-component solitons}\label{s:solitons}
First we analyse Eq. \eqref{e:model} without Raman ($\eps_f = \eps_s = 0$). Looking for soliton solutions $\psi_f$ and $\psi_s$, we use the ansatz 
\eq{\label{e:ansatz}
U_f &= \psi_f(\eta, \mu, \nu )e^{i\mu\xi},\\
U_s &= \psi_s(\eta, \mu, \nu)e^{i[2\mu- \kappa]\xi }.
}
where $\eta = \tau - \nu \xi$, $\mu$ is the nonlinear shift to the propagation constant and $\nu$ is the inverse soliton velocity. These two soliton parameters can be chosen freely, however, the relative size of $\mu$ and $\nu$ are restricted by the criteria, 
\eq{\label{e:criteria}
4r_2\mu > \nu^2, \\
4s_2[2\mu-\kappa] > [\nu-s_1]^2,
}
which ensure exponential localisation in the FF and SH component respectively \cite{Rowe2019TemporalNanowaveguides}. These existence criteria can be visualised in the $\nu$-$\mu$ plane as the areas above (below) two parabolas, in the case of anomalous (normal) dispersion, $r_2=1$, $s_2>0$ ($r_2=-1$, $s_2<0$). Both criteria must be satisfied for solitons to exist and this occurs in the region where the two parabolic areas overlap.

Substitution of our ansatz in Eq. \eqref{e:ansatz} into Eq. \eqref{e:model} results in,
\eq{\label{e:sol_mat}
\hat{S}\vec{X} = 0,
}
which we have written in matrix form for convenience where
\eq{
\hat{S} = \begin{bmatrix}
A & C^* & B & 0 \\
C & D & 0 & 0 \\
B^* & 0 & A^* & C \\
0 & 0 & C^* & D^* \\
\end{bmatrix}, \,\,\,\,\,
\vec{X} = \begin{bmatrix}
\psi_f \\
\psi_s \\
\psi_f^* \\
\psi_s^*
\end{bmatrix},
}
with matrix elements
\eq{
A =& - \mu - i\nu\partial_\eta + r_2\partial_\eta^2 \\
 &+ \alpha_f|\psi_f|^2 + \alpha_c |\psi_s|^2 , \\
B =& \psi_s/2 , \\
C =& \psi_f /2 , \\
D =& - 2\mu + \kappa + i[s_1-\nu]\partial_\eta + s_2 \partial_\eta^2 \\
 &+ \alpha_s |\psi_s|^2 + \alpha_c |\psi_f|^2.
}

Previous work has identified soliton solutions in this system of equations \cite{Buryak2002, Buryak1995OpticalNonlinearities}. Importantly localised soliton solutions are possible when the sign of dispersion in FF and SH are the same \cite{Rowe2019TemporalNanowaveguides}. We note two results of particular relevance for this work on bright temporal solitons. Firstly, in the case of anomalous dispersion many families of soliton solutions exist with only one, known as 'C-type', being stable \cite{Buryak1995OpticalNonlinearities}. Secondly, in the case of normal dispersion where $\chithree$ and $\chitwo$ nonlinearities are in opposition, only one family of soliton exists, below a certain peak power \cite{Buryak1995OpticalNonlinearities}. In both cases these solitons are non-zero in both FF and SH components which in this work we refer to as two-component solitons.

For these two-component solitons there are three regimes worth noting here. Firstly the Kerr limit when $\chithree$ nonlinearity dominates ($\psi_f|\psi_f|^2\alpha_f\gg \psi_f^*\psi_s$) and the majority of the soliton power resides in the FF component. Secondly the cascaded Kerr limit where $\chitwo$ nonlinearity is dominant ($\psi_f|\psi_f|^2\alpha_f\ll \psi_f^*\psi_s$ and $\psi_f|\psi_f|^2\alpha_c\ll \psi_f^2$) and phase mismatch is large, $|\kappa| \gg |\mu|$, with the correct sign such that $ \kappa r_2 < 0$. In this regime the interaction between SH and FF can be approximated as a cascaded Kerr term in the FF giving rise to solitons similar in form to those in the true Kerr limit and with a negligible SH component \cite{Buryak2002}. Lastly is the quadratic limit, where $\chitwo$ nonlinearity is dominant and phase mismatch is small. This regime is characterised by a more even distribution of power between the FF and SH and neither component is negligible.

\begin{figure}[b]
    \centering
    \includegraphics{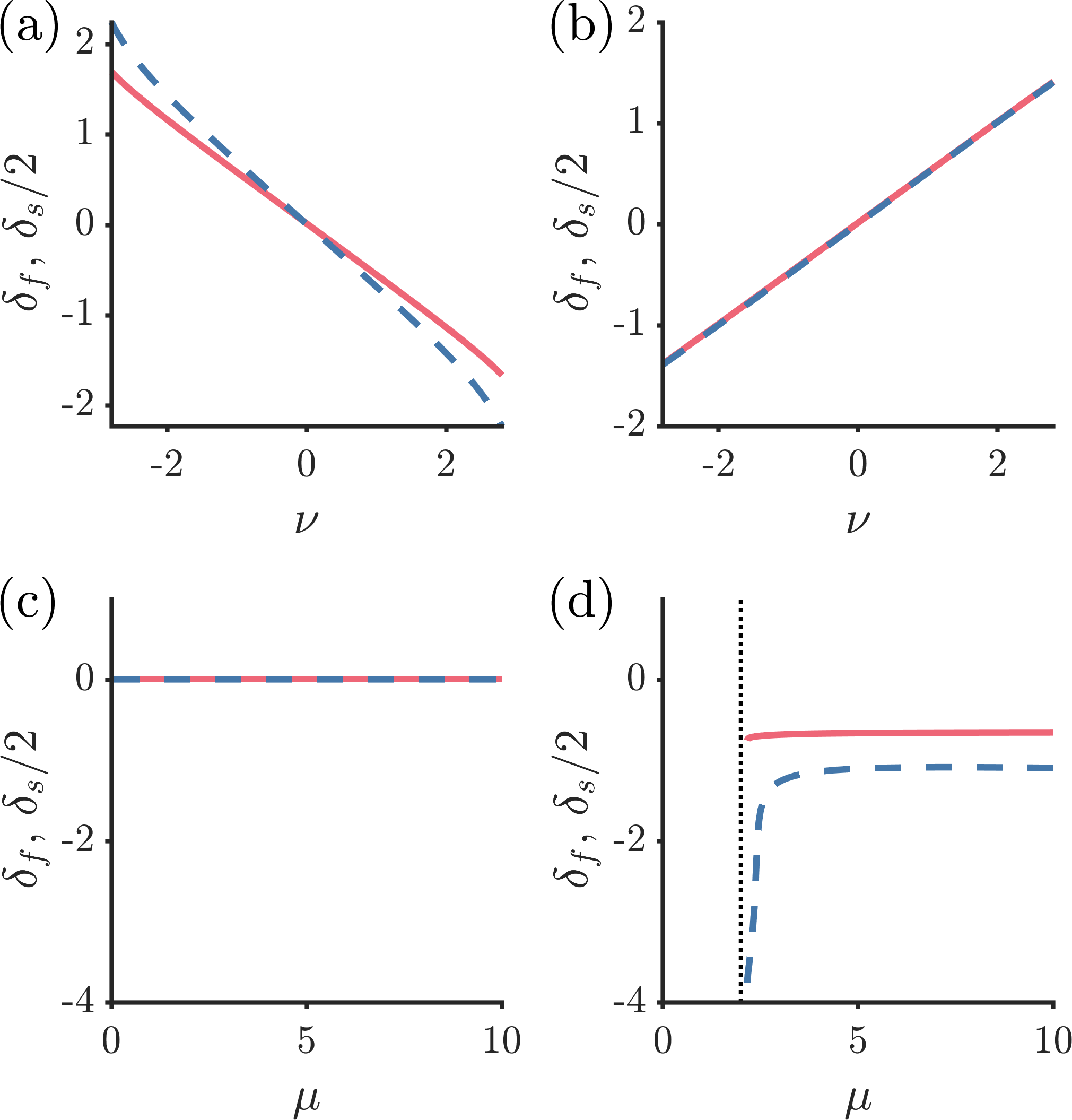}
    \caption{Frequency shift of soliton components as a function of $\nu$ and $\mu$, calculated by numerical Newton-Raphson method. $\delta_f$ and $\delta_s/2$ are plotted by solid (red) and dashed (blue) lines respectively. Panels are representative of solitons in the (a) quadratic limit with anomalous dispersion ($\mu=5$, $\kappa=0$, $s_1=0$, $r_2=1$, $s_2=1/4$) and (b) Kerr limit with normal dispersion ($\mu=-2$, $\kappa=300$, $s_1=0$, $r_2=-1$, $s_2=-1/4$ ). Panels (c) and (d) show data for solitons in the quadratic limit with anomalous dispersion ($\kappa=0$, $s_1=0$, $r_2=1$, $s_2=1/16$). Inverse soliton velocity, $\nu = 0$ and $\nu=1$ in panels (c) and (d) respectively. The black dotted line marks boundary of soliton existence for the given $\nu$ found using Eq. \eqref{e:criteria} ( $\mu = 0$ and $\mu = 2$ in panels (c) and (d) respectively). $t_0=100\rm{fs}$ throughout. }
    \label{fig:NR_freq_shift}
\end{figure}

We find two-component soliton solutions to Eq. \eqref{e:sol_mat} using a Newton-Raphson method with the known analytic quadratic soliton solution as the initial condition \cite{Buryak2002}.
Taking the Fourier Transform of these solutions we then find the frequency of the spectral peaks, $\omega_f'$ and $\omega_s'$ of the FF and SH components respectively.
We define the normalised frequency shift of each component from their reference frequencies as $\delta_f = [\omega_f' - \omega_f]t_0$ for the FF and $\delta_s$ similarly for the SH component. We note that these definitions place no constraint on the shift of the SH relative to the FF, therefore the peak frequency of the SH component may not necessarily be twice that of the FF, as shown in figure \ref{fig:NR_freq_shift} (a).
In this figure we plot $\delta_f$ and $\delta_s/2$ as a function of inverse soliton velocity $\nu$. Only where the two plotted functions coincide is the peak frequency of the SH component exactly double that of the FF. There is a clear difference between the quadratic and Kerr limits illustrated in figure \ref{fig:NR_freq_shift} (a) and (b) respectively. Considering the relative powers in each component for each regime this difference can be easily understood. In the Kerr limit the majority of the soliton power resides in the FF component and so behaviour of the SH is dominated by its interaction with the FF, thus $\delta_s$ is locked at exactly twice $\delta_f$. Conversely in the quadratic regime where the power is more evenly shared between both components, we see the spectral shifts are not dominated by a single component. The different signs of dispersion in each of figure \ref{fig:NR_freq_shift} (a) and (b) govern the sign of the gradient of $\delta_f$ and $\delta_s$. 

Figures \ref{fig:NR_freq_shift} (c) and (d) plot $\delta_f$ and $\delta_s/2$ as a function of soliton parameter $\mu$. The difference between these two panels is the inverse soliton velocity, $\nu$. The importance of this difference becomes clear when we consider the soliton existence criteria in Eq. \eqref{e:criteria}. Looking first at panel (c), where $\nu=0$, we notice that $\delta_f = \delta_s/2 =0$ for the range of $\mu$ shown. In panel (d) however, where $\nu=1$, the behaviour is not as simple. For large $\mu$ values we see that $\delta_f \neq \delta_s/2$ which is expected from panel (a). For smaller $\mu$ values however we see that $\delta_s$ shifts rapidly as $\mu$ approaches the boundary of soliton existence without much change in $\delta_f$. Unlike in Kerr solitons there is no simple relationship between shifts in $\nu$ and $\mu$ and the spectral shifts in two-component solitons. It is a complicated dependence reflecting the balance of power between and dispersion of each component.

\section{Raman shift theory}
Now we consider the effects of a small but non-zero Raman response ($\eps_f=\eps_s \neq 0$). To accommodate this change our ansatz in Eq. \eqref{e:ansatz} becomes,
\eq{
U_f &= [\psi_f(\eta,\mu,\nu) + a_f(\xi, \eta) ]e^{i\phi},\\
U_s &= [\psi_s(\eta,\mu,\nu)+ a_s(\xi, \eta) ]e^{i2\phi-i\kappa\xi },
}
where we now allow $\mu$ and $\nu$ to slowly vary with $\xi$. Our transverse coordinate, $\eta = \tau - \int \nu \dd \xi$, keeps pace with the soliton as its inverse velocity $\nu$ changes. $\phi = \int \mu \dd \xi$ allowing for the phase of the soliton to evolve as the propagation constant $\mu$ changes with $\xi$. We introduce the small terms $a_f \ll \psi_f$ and $a_s \ll \psi_s$ to allow for corrections to the soliton profile and radiation from the solitons. Now the soliton parameters $\nu$ and $\mu$ can vary we expect that the position of the soliton will move on the $\nu$-$\mu$ plane where without Raman the soliton was previously a static point. The trajectory that the soliton takes and how it interacts with the existence criteria in Eq. \eqref{e:criteria} will be discussed in Sec. \ref{s:sim_sol_instability}.

Writing our governing equation in matrix form we find,
\eq{\label{e:pert_Eq.}
\hat{S}\vec{X} + \hat{P}\vec{X} + \hat{J}\vec{a} = \vec{\eps},
}
where $\hat{S}$ and $\vec{X}$ are defined as before and
\eq{
&\hat{P} = i \begin{bmatrix}
G & 0 & 0 & 0 \\
0 & G & 0 & 0 \\
0 & 0 & -G^* & 0 \\
0 & 0 & 0 & -G^* \\
\end{bmatrix}, \,\,\,\,\,
&\vec{a} = \begin{bmatrix}
a_f\\
a_s\\
a_f^*\\
a_s^*\\
\end{bmatrix},\\
&\hat{J} = \begin{bmatrix}
M & N & Q & R \\
N^* & T & R & W \\
Q^* & R^* & M^* & N^* \\
R^* & W^* & N & T^* \\
\end{bmatrix}, \,\,\,\,\,
&\vec{\eps} = \begin{bmatrix}
\tilde{\eps}_f\\
\tilde{\eps}_s \\
\tilde{\eps}_f^* \\
\tilde{\eps}_s^* \\
\end{bmatrix},
}
with the matrix elements
\eq{
G &= [\partial_\xi \mu] \partial_\mu + [\partial_\xi \nu ]\partial_\nu , \\
M &= -\mu -i\nu\partial_\eta  + r_2\partial_\eta^2+ 2\alpha_f |\psi_f|^2 + \alpha_c |\psi_s|^2, \\
N &= \psi_f^* + \alpha_c \psi_s^* \psi_f,\\
Q &= \psi_s + \alpha_f \psi_f^2,\\
R &= \alpha_c \psi_s \psi_f,\\
T &= -[2\mu - \kappa] + i[s_1-\nu] \partial_\eta + s_2 \partial_\eta^2 \\&\qquad\qquad\qquad\qquad\qquad+ 2\alpha_s |\psi_s|^2 + \alpha_c |\psi_f|^2,\\
W &= \alpha_s \psi_s^2
}
We have kept terms up to the order of $\vec{a}$, smaller terms have been neglected. As $\vec{\eps}$ is also a small quantity, terms proportional to $\vec{\eps}\vec{a}$ are very small and so are neglected. We also introduce the notation $\tilde{\eps}_f = \eps_f(\psi_f, \psi_s)$ and $\tilde{\eps}_s = \eps_s(\psi_f, \psi_s)$.

Recognising from before that $\hat{S}\vec{X} = 0$, we reduce the left hand side in Eq. \eqref{e:pert_Eq.} to two terms. To make further progress we project Eq. \eqref{e:pert_Eq.} onto the zero eigen-vectors of $\hat{J}^{\dagger} = \hat{J}$ ( as $\hat{J}$ is self-adjoint). So Eq. \eqref{e:pert_Eq.} becomes
\eq{
(\vec{v}_n,\hat{P}\vec{X}) + (\vec{v}_n,\hat{J}\vec{a}) = (\vec{v}_n, \vec{\tilde{\eps}}),
}
where the term $\hat{J}\vec{a}$ vanishes by construction. $n$ is either $1$ or $2$ such that $\vec{v}_n$ is one of the two zero-eigenvectors of $\hat{J}$,
\eq{
\vec{v}_1 = \partial_\eta \vec{X}, \qquad
\vec{v}_2 = \begin{bmatrix}  \psi_f \\ 2\psi_s\\ -\psi_f^*\\ -2\psi_s^* \end{bmatrix}.
}
 The results of projecting Eq. \eqref{e:pert_Eq.} on to $\vec{v}_1$ and $\vec{v}_2$ are respectively,
\eq{\label{e:Im_Re}
\Im \int_{-\infty}^{\infty}& [\partial_\eta \psi_f^*] [G \psi_f] +  [\partial_\eta \psi_s^*] [G \psi_s]  \dd \eta \\&= -\Re \int_{-\infty}^{\infty}  [\partial_\eta \psi_f^*] \tilde{\eps}_f  +[\partial_\eta \psi_s^*] \tilde{\eps}_s   \dd \eta, \\
\Re \int_{-\infty}^{\infty}&  \psi_f^* [G \psi_f] +  2 \psi_s^* [G \psi_s]  \dd \eta \\&= \Im \int_{-\infty}^{\infty}  \psi_f^* \tilde{\eps}_f  + 2\psi_s^* \tilde{\eps}_s   \dd \eta.
}
Using these two equations we find the shift of our two soliton parameters $\mu$ and $\nu$ to be,
\eq{\label{e:mu_nu_shift}
\dot{\nu} = \diffp{\nu}{\xi} =& \frac{C_2 A_1 - C_1 A_2}{B_2 A_1 - B_1 A_2}, \\ 
\dot{\mu} = \diffp{\mu}{\xi} =& \frac{C_1 B_2 - C_2 B_1}{B_2 A_1 - B_1 A_2},
}
where we introduce the parameters, 
\eq{
A_1 =& \Im   \int_{-\infty}^{\infty}   [\partial_\eta \psi_f^* ] [\partial_\mu \psi_f]  + [\partial_\eta \psi_s^* ] [\partial_\mu  \psi_s]  \, \dd \eta ,\\
B_1 =& \Im   \int_{-\infty}^{\infty}   [\partial_\eta \psi_f^*]  [\partial_\nu   \psi_f]  + [\partial_\eta \psi_s^* ][\partial_\nu  \psi_s]  \, \dd \eta ,\\
C_1 =& \Im   \int_{-\infty}^{\infty}   [\partial_\eta \psi_f^* ] \tilde{\eps}_f  + [\partial_\eta \psi_s^* ] \tilde{\eps}_s \, \dd \eta ,\\
A_2 =& \Re   \int_{-\infty}^{\infty}   \psi_f^* [\partial_\mu \psi_f]  + 2 \psi_s^* [\partial_\mu \psi_s]  \, \dd \eta ,\\
B_2 =& \Re   \int_{-\infty}^{\infty}   \psi_f^* [\partial_\nu \psi_f]  + 2 \psi_s^* [\partial_\nu  \psi_s]  \, \dd \eta ,\\
C_2 =& \Re   \int_{-\infty}^{\infty}   \psi_f^* \tilde{\eps}_f    + 2 \psi_s^* \tilde{\eps}_s   \, \dd \eta .\\
}
We point out that these parameters are themselves all $\mu$ and $\nu$ dependent through $\psi_f$ and $\psi_s$.
From Eq. \eqref{e:mu_nu_shift} we integrate over $\xi$ using numerical soliton profiles $\psi_f(\mu, \nu)$ and $\psi_s(\mu, \nu)$ to find the inverse soliton velocity, $\nu(\xi)$ and the soliton propagation constant $\mu(\xi)$. As a result of the $\nu$ and $\mu$ dependence of $\psi_f$ and $\psi_s$, accurate calculation of these shifts requires repeated evaluation of Eq. \eqref{e:mu_nu_shift} with $\psi_f$ and $\psi_s$ being updated on each iteration for the current shifted $\mu$ and $\nu$ values.

Once $\nu(\xi)$ and $\mu(\xi)$ are known we can then calculate the temporal shift of the soliton 
\eq{\label{e:time_shift}
\tau_s(\xi) = \int_0^\xi \nu(\xi') \dd \xi'.
}
and the shift in the peak frequency of each soliton component,
\eq{\label{e:freq_shift}
\dot{\delta_k} = \diffp{\delta_k}{\xi} = [\partial_\nu \delta_k] \dot{\nu} + [\partial_\mu \delta_k] \dot{\mu},
}
where $k$ is either $f$ or $s$ for FF or SH respectively. Both $\partial_\nu \delta_k$ and $\partial_\mu \delta_k$ can in general be computed numerically for each soliton component as described in Sec. \ref{s:solitons}.

\subsection{Kerr Limit} 

It is useful to consider the Kerr limit and observe how our results in Eq. \eqref{e:mu_nu_shift} and $\eqref{e:freq_shift}$ converge to the well-known results for pure Kerr solitons. In the Kerr limit we neglect $U_s$ and require anomalous dispersion ( $r_2=1$). With these changes our model in Eq. \eqref{e:model} becomes the nonlinear Schr{\"o}dinger equation (NLS),
\eq{
i\partial_\xi U_f + \partial_\tau^2 U_f +  \alpha_f U_f |U_f|^2 = \eps_f(U_f,0),
}
which is invariant under the Galilean transform \cite{Kivshar2003ChapterSolitons}
\eq{\label{e:Gal_inv}
U_f &\rightarrow U_f \exp(i\frac{v}{2}\eta + i\frac{v^2}{4}\xi),\\
\tau &\rightarrow \tau - \nu \xi = \eta.
}
The NLS has known solutions for solitons of velocity $\nu$ \cite{Kivshar2003ChapterSolitons},
\eq{
U_f(\eta) &= \psi_0 \exp(i\mu\xi + i \frac{\nu}{2}\eta)
}
where we have defined the real soliton profile \cite{Agrawal2013NonlinearOptics},
\eq{\label{e:psi_0}
\psi_0 &= \sqrt{\frac{2q}{\alpha_f}}\sech(\sqrt{q}\eta),
}
where $q = \mu - \nu^2/4$. From our previous ansatz, Eq. \eqref{e:ansatz}, we can see that $\psi_f = \psi_0 \exp(i\nu\eta/2)$. Making this substitution into our predictions of Raman shift in Eq. \eqref{e:Im_Re} and remembering to neglect $\psi_s$ terms we find,
\eq{\label{e:Im_kerr_step1}
 \int_{-\infty}^{\infty}& \frac{\eta}{2}   \psi_0 \dot{\nu}[\partial_\eta \psi_0]  - \frac{\nu}{2}\psi_0\big[ \dot{\mu} [\partial_\mu \psi_0] + \dot{\nu} [\partial_\nu \psi_0] \big]   \dd \eta \\& \,\,\,\,\,\,\,\,\,\, = \int_{-\infty}^{\infty} -2\alpha_f\tau_R \psi_0^2 [\partial_\eta\psi_0]^2 \dd \eta, 
 }
 and
 \eq{\label{e:Re_kerr_step1}
 \int_{-\infty}^{\infty}& \psi_0\big[\dot{\mu}[\partial_\mu \psi_0 ] + \dot{\nu} [\partial_\nu \psi_0] \big] \dd \eta = 0
}
where we have kept only the terms that contribute after taking real and imaginary parts. We use the approximate, small bandwidth form of the Raman response in silica $\tilde{\eps}_f = \psi_f \tau_R \partial_\eta[\alpha_f |\psi_f|^2]$ where $\tau_R = f_R \int \tau R(\tau) \dd \tau = 0.0073$ \cite{Skryabin2005TheoryFibers}. By substitution of Eq. \eqref{e:Re_kerr_step1} and \eqref{e:psi_0} into Eq. \eqref{e:Im_kerr_step1} and integrating we restore the known result for the Kerr limit in silica \cite{Agrawal2013NonlinearOptics, Skryabin2010TheoryWaves},
\eq{\label{e:nu_dot_kerr}
\dot{\nu} = \frac{32}{15} \tau_R q^2,
}
where we have used the integrals
\eq{
\int_{-\infty}^{\infty} x \sech^2(x) \tanh(x) \dd x  = 1,\\
\int_{-\infty}^{\infty} \sech^4(x) \tanh^2(x) \dd x  = \frac{4}{15}.
}
Similarly integration of Eq. \eqref{e:Re_kerr_step1} gives,
\eq{
\dot{\mu} = \frac{\nu}{2} \dot{\nu},
}
and therefore we restore
\eq{\label{e:mu_xi_kerr}
\mu(\xi) = q + \frac{[\nu(\xi)]^2}{4},
}
as expected from the Galilean transform in Eq. \eqref{e:Gal_inv}.

From Eq. \eqref{e:Gal_inv} we know that a change in inverse soliton velocity, $\nu$, results in a subsequent change to the solitons frequency $\delta_f = -\nu/2$. Using this we can find the rate of frequency shift,
\eq{\label{e:delta_dot}
\dot{\delta_f} = [\partial_\nu \delta_f] \dot{\nu}= -\frac{16}{15}\tau_R q^2,
}
which also agrees with known results for the Kerr limit in silica\cite{Agrawal2013NonlinearOptics, Skryabin2010TheoryWaves}.

\begin{figure}
    \centering
    \includegraphics[]{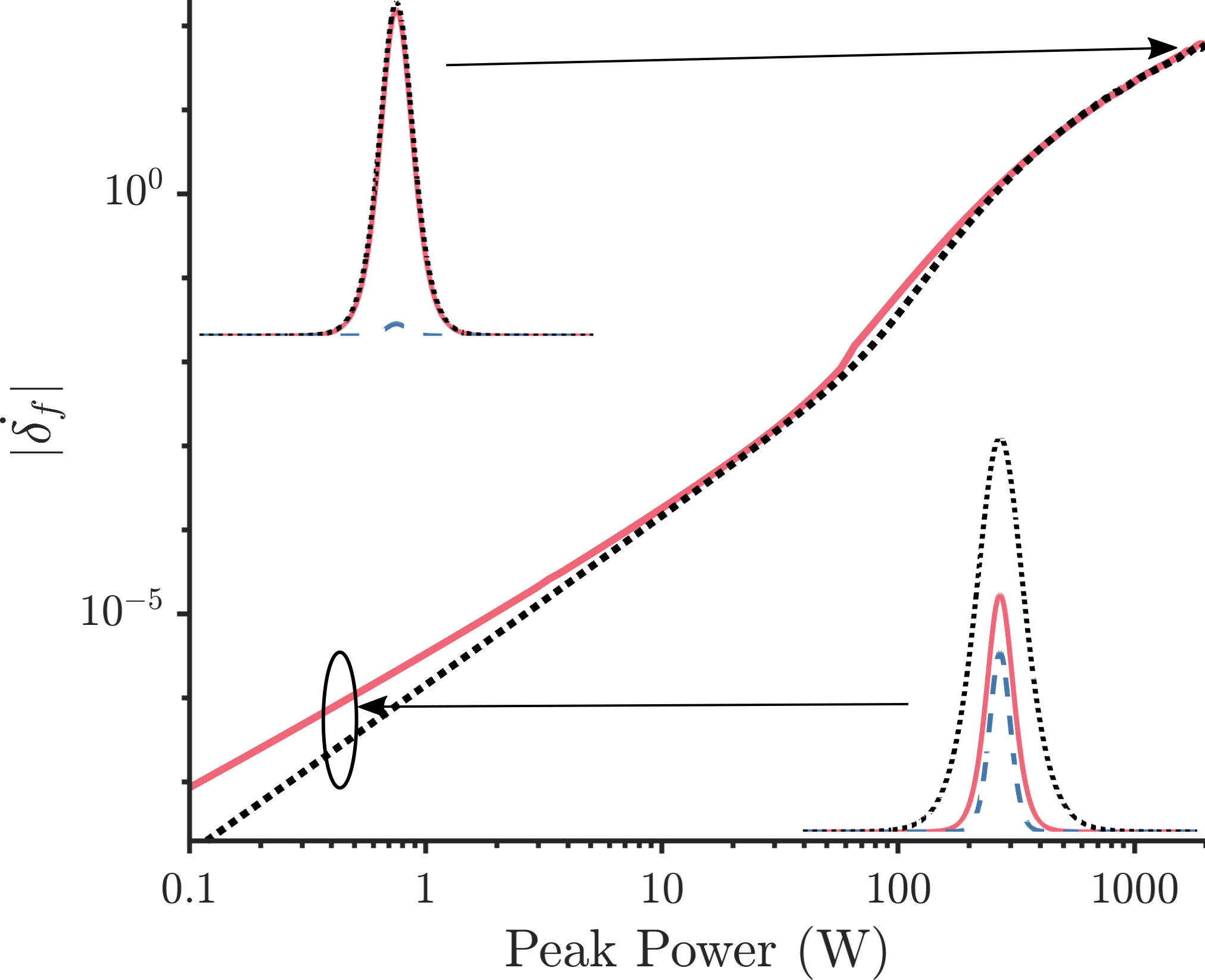}
    \caption{ Rate of frequency shift, $\dot{\delta}_f$, predicted by Eq. \eqref{e:freq_shift} for two-component solitons and Kerr solitons in LN nano-waveguides; shown as solid (red) and dotted (black) curves respectively (for $\nu=0$). For the two-component solitons we plot the peak power as $|A_f|^2 + |A_s|^2$. Insets compare soliton power profiles at high and low powers. FF and SH of the two-component solitons are plotted as solid (red) and dashed (blue) curves respectively. Kerr soliton power profile shown as dotted (black) curve with the same total peak power. We found that $\dot{\delta}_s/2$ very closely followed the same trend as $\dot{\delta}_f$ for the two-component soliton in this case.}
    \label{fig:freq_shift}
\end{figure}

For the general case, where the Raman response is not simply linear over the bandwidth of the soliton (as in LN) or where the SH component is significant, the Raman shifts can be calculated numerically. Doing this for LN, we find that in the Kerr limit the FF component of the two-component soliton shifts in frequency at the same rate as a Kerr soliton of the same peak power. In the quadratic limit, when the soliton has a significant SH component, we find that the FF frequency shift differs significantly from that expected for a Kerr soliton of the same peak power. This trend is illustrated in figure \ref{fig:freq_shift}. We compare solitons of the same peak power as Eq. \eqref{e:delta_dot} shows us the simple relationship between $\dot{\delta_f}$ and $q$ (which from Eq. \eqref{e:psi_0} is proportional to soliton peak power) in the Kerr limit.

\section{Simulations}

In order to assess their validity outside of the Kerr limit and linear Raman response regime, we have compared our predictions in Eq. \eqref{e:time_shift} and \eqref{e:freq_shift} with numerical simulations of Eq. \eqref{e:model} using the split-step Fourier method. We use numerical soliton solutions as the initial inputs for both the simulations and predictions. In these comparisons we chose dispersion and soliton parameters that are feasible in LN nano-waveguides at reasonable input powers.
\begin{figure*}
    \centering
    \includegraphics[]{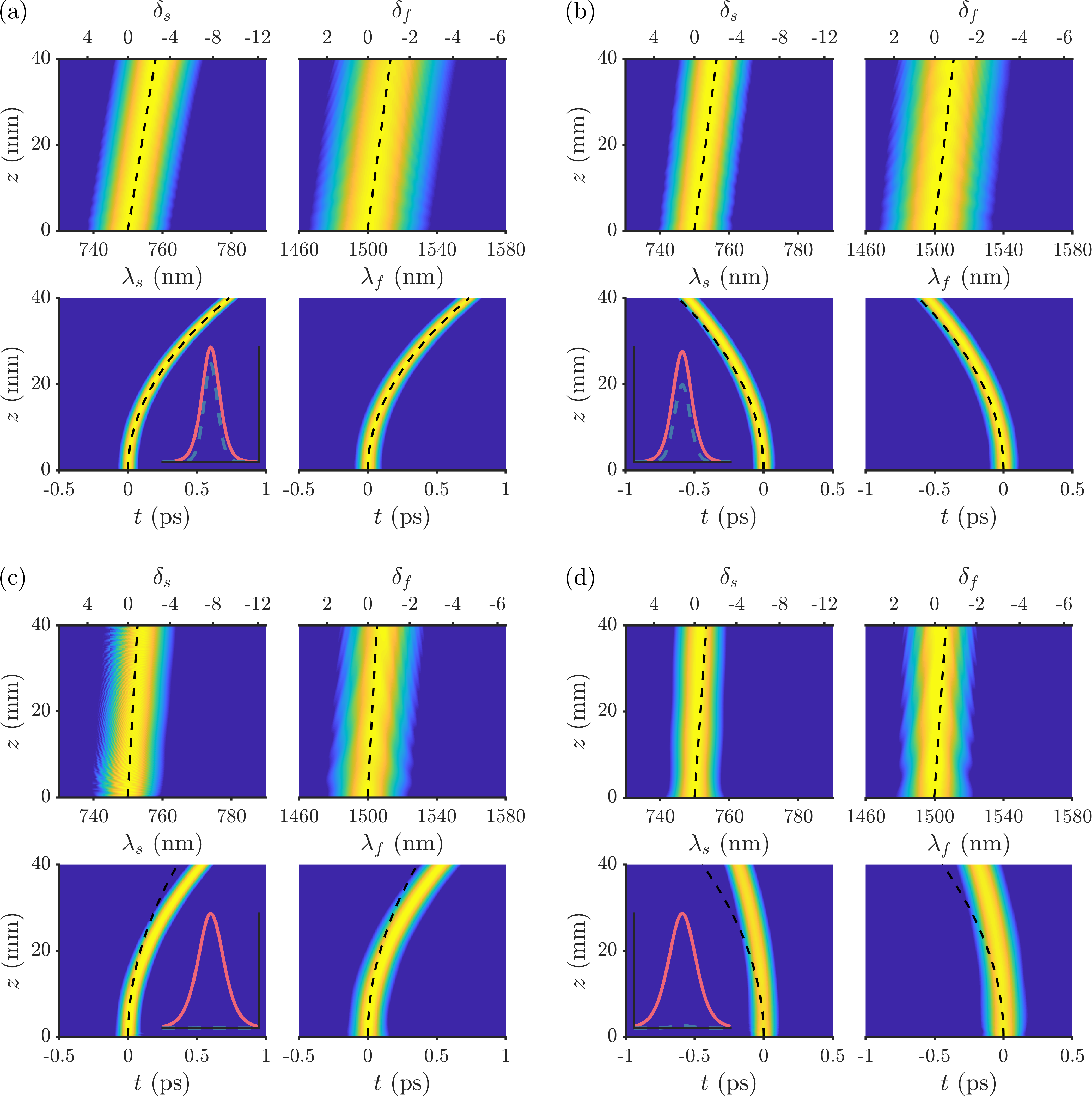}
    \caption{ Simulated soliton propagation in time and frequency domains shown in the upper and lower plots in each panel respectively. The FF and SH are shown on the right and left sides of each panel respectively. Dashed lines show time and frequency shifts predicted by Eq. \eqref{e:time_shift} and \eqref{e:freq_shift}. Insets show the relative input soliton power plotted in the time domain for FF (solid red) and SH (dashed blue). Panels (a) and (b) show solitons in the quadratic limit with anomalous and normal dispersion respectively. Panels (c) and (d) show solitons in the cascaded Kerr limit with anomalous and normal dispersion respectively. Parameters used in each panel are: (a) $\mu=5$, $\nu=0$, $\kappa=0$, $s_1=0$, $r_2=1$, $s_2=1/4$. (b) $\mu=-5$, $\nu=0$, $\kappa=0$, $s_1=0$, $r_2=-1$, $s_2=-1/4$. (c) $\mu=2$, $\nu=0$, $\kappa=-1000$, $s_1=10$, $r_2=1$, $s_2=1/4$. (d) $\mu=-2$, $\nu=0$, $\kappa=300$, $s_1=0$, $r_2=-1$, $s_2=-1/4$. For all of these simulations we used an input FF wavelength of $1500 \rm{nm}$ ( $\omega_f = 200\rm{THz}$), characteristic time scale, $t_0 = 100\rm{fs}$ and $\alpha_f = 0.0007$, $\alpha_s=0.008$, $\alpha_c=0.005$.
    }
    \label{fig:sim_prop}
\end{figure*}

We present four typical examples of such comparisons in figure \ref{fig:sim_prop} which are representative of two soliton limits each in two dispersion regimes. In two of these examples dispersion is anomalous and phase mismatch of $\kappa =0$ and $\kappa = 300$ provide both quadratic and cascaded Kerr limits respectively. The other two examples explore the normal dispersion regime again with quadratic and cascaded Kerr limits represented. We ensure the Raman term is a perturbation to the soliton equation by operating in the quadratic and cascaded Kerr regimes. This keeps all Kerr terms, including Raman, relatively small. We also used an initial inverse soliton velocity, $\nu=0$ such that the input pulse was not shifted from the reference frequencies initially(i.e. $\delta_f=\delta_s = 0$ at $\xi=0$).

Figures \ref{fig:sim_prop} (a) and (b) show remarkable similarity between prediction and simulation for Raman shifting quadratic two-component solitons under anomalous and normal dispersion respectively. These examples are particularly important as they validate our predictions outside of the Kerr limit, in regimes where the SH soliton component is similar in power to that of the FF. In figures \ref{fig:sim_prop} (c) and (d) the phase mismatch, $\kappa$ is set such that the solitons are in the cascaded Kerr regime with very small SH components. In these examples we see a qualitative agreement between our predictions and simulations. The reason some predictions do not match the simulations more closely is largely down to spectral cut-off of the soliton due to the sharp Raman peak in LN \cite{Gorbach2008SolitonFibers}. The sharp Raman peaks in LN are clear when compared with silica in figure \ref{fig:Raman_response}. This has been seen before in solitons in air-core fibres where a significant portion of the initial pulse propagates as non-solitonic radiation \cite{Gorbach2008SolitonFibers}. The soliton propagating in the simulation is therefore not necessarily the same as the input soliton for which the predictions are made.

We point out that although the Raman frequency shift is always towards lower frequencies as expected, the temporal shift of the soliton is dependant on the sign of dispersion. Under anomalous dispersion lower frequencies travel slower and so as the soliton shifts to lower frequencies it slows down moving in the positive time direction (arriving late). Under normal dispersion the opposite is true (and the soliton arrives early).

\subsection{ Soliton Instabilities }\label{s:sim_sol_instability}

We now consider the soliton trajectory in the $\nu$-$\mu$ plane, predicted by Eq. \eqref{e:mu_nu_shift}, in relation to the soliton existence criteria given in Eq. \eqref{e:criteria}. Doing so prompts us to ask two questions; does the trajectory of the soliton approach the boundary of soliton existence and if so, what happens to the soliton when it does?

Considering Kerr solitons, only one existence criterion is relevant which is similar to that of the FF with anomalous dispersion ($r_2=1$). The criterion for soliton existence is therefore $4\mu > \nu^2 $. We note that Kerr solitons follow the trajectory described by Eq. \eqref{e:mu_xi_kerr}, which we notice is always tangential to the existence boundary, $\mu=\nu^2/4$. This is not a coincidence and is due to the simple relation between the shift in soliton frequency and inverse velocity, $\nu$, given by the Galilean transform in Eq. \eqref{e:Gal_inv}. A Raman shifting Kerr soliton therefore does not approach the boundary of soliton existence.

In the case of two-component solitons however, the relation between frequency and inverse velocity shifts is not simple. There are also two existence criteria that must both be satisfied and which in general do not coincide. It is therefore unclear whether a Raman shifting two-component soliton will approach the boundary of soliton existence.

\begin{figure*}
    \centering
    \includegraphics[]{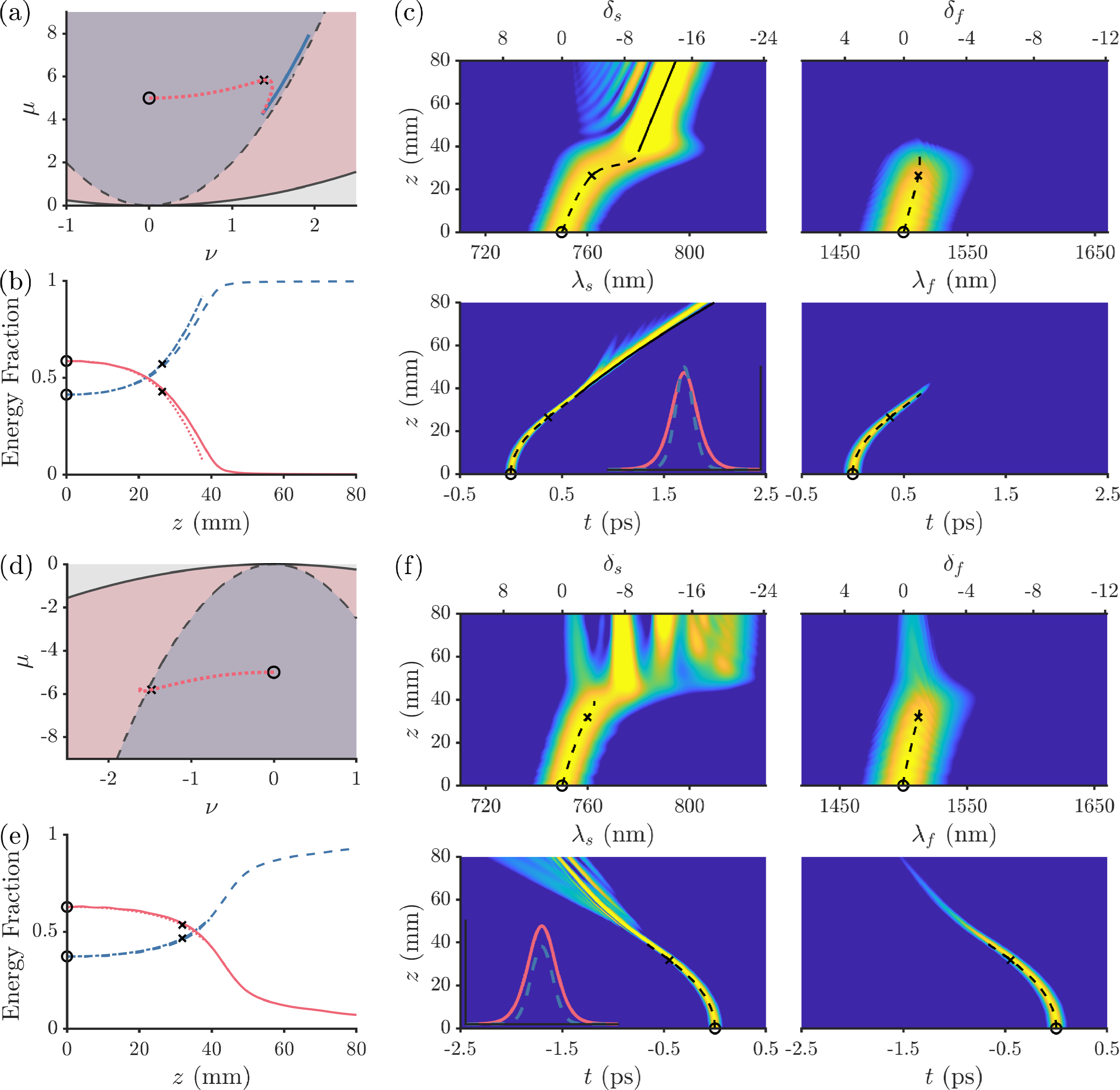}
    \caption{ Prediction and simulation of soliton instabilities. Panels (a)-(c) show an example under anomalous dispersion with the same parameters as figure \ref{fig:sim_prop} (a) except with $s_2 = 1/16$. Panels (d)-(f) show an example under normal dispersion with parameters as in figure \ref{fig:sim_prop} (b) but with $s_2=-1/20$. Soliton existence analysis is shown in panels (a) and (d), shaded regions mark where each soliton existence criterion is met. The two-component soliton trajectory in the $\nu$-$\mu$ plane predicted by Eq. \eqref{e:mu_nu_shift} is included as the dotted (red) curve. Solid (blue) curve in panel (a) shows the predicted trajectory of the Kerr soliton in the SH after the two-component soliton becomes unstable. Panels (b) and (e) plot fractional energy in each component of the soliton as a function of propagation distance. Data from the simulated soliton propagation are plotted as solid (red) and dashed (blue) curves for the FF and SH respectively. Fractional energies predicted by our analysis for FF and SH shown as dotted (red) and dot-dashed (blue) lines respectively, calculated from numerical solitons for the predicted $\nu$ and $\mu$ values. Panels (c) and (f) plot simulated propagation of the solitons in each example. Top and bottom rows of each panel show the spectral and temporal evolution respectively, SH and FF are shown on the left and right columns respectively. Trajectory of the soliton predicted by Eq. \eqref{e:freq_shift} and \eqref{e:time_shift} are plotted as dashed (black) curves in the spectral and temporal plots respectively. Solid (black) curves in panel (c) plot predictions for a Kerr soliton in the SH after the two-component soliton becomes unstable. Insets plot relative input soliton power in the time domain for FF (solid red) and SH (dashed blue). Circles mark the initial soliton position, crosses mark a particular point of interest in each (a)-(c) and (d)-(f). 
    }
    \label{fig:sol_instability}
\end{figure*}

Here we give two examples that demonstrate that two-component solitons can approach the boundary of existence. In these examples we use systems where the group velocity dispersion in one component is significantly smaller than the other (e.g. $s_2 \ll r_2$). These parameters are chosen as a simple means of narrowing the region of soliton existence, as shown in figures \ref{fig:sol_instability} (a) and (d), thus reducing the shift required to approach the existence boundary. Tracking the soliton as it shifts in $\mu$ and $\nu$ we find that initially the soliton shifts parallel to the FF localisation boundary, and therefore approaches the boundary existence which in these cases is the boundary of SH localisation. Close to this boundary we find our predictions in Eq. \eqref{e:mu_nu_shift} break down as calculation of updated soliton solutions becomes more difficult. After this we rely on our simulations to see what happens.

In the example in figure \ref{fig:sol_instability} (a)-(c), dispersion is anomalous and we find that the soliton appears to recoil \cite{Skryabin2003SolitonFibers} to avoid the existence boundary, as shown in panel (a). At the point marked with a cross in figures \ref{fig:sol_instability} (a)-(c) the soliton abruptly changes trajectory, shifting down in $\mu$ and very little in $\nu$, still approaching the existence boundary. In figure \ref{fig:sol_instability} (c) we see that this recoil in the $\nu$-$\mu$ plane is accompanied by a rapid frequency shift of the SH component, shown in the simulated spectrum and replicated by remarkably closely our predictions. We note that this frequency shift is similar to that shown in figure \ref{fig:freq_shift} (d) where $\delta_s$ rapidly shifts as $\mu$ reduces and the soliton approaches the existence boundary. In this example we see the majority of the soliton energy is transferred to the SH as shown clearly in figure \ref{fig:sol_instability} (b) where again our predictions closely follow the initial stages of the simulation. After the two-component soliton is destabilised a purely Kerr soliton appears to form in the SH with minimal energy in the FF component. Comparing the predicted spectral and temporal shifts for a SH Kerr soliton of the same peak power with the simulations in figure \ref{fig:sol_instability} (c) we see they match very closely. In figure \ref{fig:sol_instability} (a) we plot the trajectory on the $\nu$-$\mu$ plane predicted for this Kerr soliton which we find is tangential to the boundary of existence as expected. We therefore conclude that,  the outcome of the two-component soliton instability is the birth of a Kerr soliton in the SH. 

Under normal dispersion the soliton initially behaves similarly, shifting towards the boundary of SH existence as shown in figure \ref{fig:sol_instability} (d). In this case however we do not see any recoil of the soliton in the $\nu$-$\mu$ plane at the position marked by the cross. As there is no recoil in the $\nu$-$\mu$ plane, we predict no rapid spectral shift. This is confirmed by simulations in figure \ref{fig:sol_instability} (f) where the rapid frequency shift seen in the SH occurs well after our predictions stop and therefore after the soliton has shifted out its region of existence. Once again we see a large transfer of energy from the FF to the SH which is presented in figure \ref{fig:sol_instability} (e), although not as complete as in the anomalous dispersion case. We can see from the simulation in figure \ref{fig:sol_instability} (f) that the SH and remaining FF components disperse after the instability producing a broad spectrum in the SH. We do not expect a Kerr soliton in the SH as this is not possible under normal dispersion.

We have also run predictions and simulations of systems where $r_2 \ll s_2$ and find similar results. The major difference in these cases is that the power is transferred to the FF rather than the SH during the instability development. 

\section{Conclusion}
In this work have developed theory of Raman induced self-frequency shift in two-component solitons accounting for both $\chitwo$ and $\chithree$ nonlinearities. Our analysis predicts the acceleration and accompanying frequency shift of each soliton component. Expressions for the temporal and propagation constant shifts have also been presented. Known results for purely Kerr solitons have been reproduced by analysing two-component solitons in the Kerr limit and we show how our results differ outside this limit. Our predictions are validated by comparison with numerical simulation of soliton propagation in LN nano-waveguides which show good qualitative agreement in every case and remarkable quantitative agreement in some cases. 

We have proposed that two-component Raman solitons may approach the boundary of their own existence which we shown does not happen in Kerr solitons. We have described a few scenarios of the unstable dynamics of the two-component solitons induced by the Raman effect shifting the soliton frequency  towards the boundary of its existence. Our theory predicts the initial stage of this process and matches remarkably closely with simulations. 

The predictions made here can be applied to future work on ultrashort pulses in systems with $\chitwo$ and $\chithree$ nonlinearity. In particular, theory of two-component Raman solitons may prove useful in the engineering and understanding of new sources of supercontinuum generation and frequency conversion. These results also show the potential of this system to produce novel soliton dynamics.
\\

All data supporting this study are openly available from the University of Bath Research Data Archive \cite{RoweDatasetHttps://doi.org/10.15125/BATH-00808}.

\section{Acknowledgement}
WRR acknowledges funding and support from the U.K. Engineering and Physical Sciences Research Council (EPSRC) Centre for Doctoral Training in Condensed Matter Physics (CDTCMP), Grant No. EP/L015544/1.

\bibliography{references}

\begin{thebibliography}{25}%
\makeatletter
\providecommand \@ifxundefined [1]{%
 \@ifx{#1\undefined}
}%
\providecommand \@ifnum [1]{%
 \ifnum #1\expandafter \@firstoftwo
 \else \expandafter \@secondoftwo
 \fi
}%
\providecommand \@ifx [1]{%
 \ifx #1\expandafter \@firstoftwo
 \else \expandafter \@secondoftwo
 \fi
}%
\providecommand \natexlab [1]{#1}%
\providecommand \enquote  [1]{``#1''}%
\providecommand \bibnamefont  [1]{#1}%
\providecommand \bibfnamefont [1]{#1}%
\providecommand \citenamefont [1]{#1}%
\providecommand \href@noop [0]{\@secondoftwo}%
\providecommand \href [0]{\begingroup \@sanitize@url \@href}%
\providecommand \@href[1]{\@@startlink{#1}\@@href}%
\providecommand \@@href[1]{\endgroup#1\@@endlink}%
\providecommand \@sanitize@url [0]{\catcode `\\12\catcode `\$12\catcode
  `\&12\catcode `\#12\catcode `\^12\catcode `\_12\catcode `\%12\relax}%
\providecommand \@@startlink[1]{}%
\providecommand \@@endlink[0]{}%
\providecommand \url  [0]{\begingroup\@sanitize@url \@url }%
\providecommand \@url [1]{\endgroup\@href {#1}{\urlprefix }}%
\providecommand \urlprefix  [0]{URL }%
\providecommand \Eprint [0]{\href }%
\providecommand \doibase [0]{http://dx.doi.org/}%
\providecommand \selectlanguage [0]{\@gobble}%
\providecommand \bibinfo  [0]{\@secondoftwo}%
\providecommand \bibfield  [0]{\@secondoftwo}%
\providecommand \translation [1]{[#1]}%
\providecommand \BibitemOpen [0]{}%
\providecommand \bibitemStop [0]{}%
\providecommand \bibitemNoStop [0]{.\EOS\space}%
\providecommand \EOS [0]{\spacefactor3000\relax}%
\providecommand \BibitemShut  [1]{\csname bibitem#1\endcsname}%
\let\auto@bib@innerbib\@empty
\bibitem [{\citenamefont {Agrawal}(2013)}]{Agrawal2013NonlinearOptics}%
  \BibitemOpen
  \bibfield  {author} {\bibinfo {author} {\bibfnamefont {G.~P.}\ \bibnamefont
  {Agrawal}},\ }\href
  {https://search.ebscohost.com/login.aspx?direct=true&db=nlebk&AN=486042&site=ehost-live}
  {\emph {\bibinfo {title} {{Nonlinear Fiber Optics}}}},\ \bibinfo {series}
  {Optics and Photonics}, Vol.\ \bibinfo {volume} {5th ed}\ (\bibinfo
  {publisher} {Academic Press},\ \bibinfo {address} {Burlington},\ \bibinfo
  {year} {2013})\BibitemShut {NoStop}%
\bibitem [{\citenamefont {Kivshar}\ \emph
  {et~al.}(2003{\natexlab{a}})\citenamefont {Kivshar}, \citenamefont
  {Agrawal},\ and\ \citenamefont {Agrawal}}]{Kivshar2003ChapterSolitonsb}%
  \BibitemOpen
  \bibfield  {author} {\bibinfo {author} {\bibfnamefont {Y.~S.}\ \bibnamefont
  {Kivshar}}, \bibinfo {author} {\bibfnamefont {G.~P.}\ \bibnamefont
  {Agrawal}}, \ and\ \bibinfo {author} {\bibfnamefont {G.~P.}\ \bibnamefont
  {Agrawal}},\ }in\ \href {\doibase
  https://doi.org/10.1016/B978-012410590-4/50003-6} {\emph {\bibinfo
  {booktitle} {Optical Solitons}}},\ \bibinfo {editor} {edited by\ \bibinfo
  {editor} {\bibfnamefont {Y.~S.}\ \bibnamefont {Kivshar}}, \bibinfo {editor}
  {\bibfnamefont {G.~P.}\ \bibnamefont {Agrawal}}, \ and\ \bibinfo {editor}
  {\bibfnamefont {G.~P.}\ \bibnamefont {Agrawal}}}\ (\bibinfo  {publisher}
  {Academic Press},\ \bibinfo {address} {Burlington},\ \bibinfo {year} {2003})\
  pp.\ \bibinfo {pages} {63--103}\BibitemShut {NoStop}%
\bibitem [{\citenamefont {Dudley}\ and\ \citenamefont
  {Taylor}(2010)}]{Dudley2010SupercontinuumFibers}%
  \BibitemOpen
  \bibfield  {author} {\bibinfo {author} {\bibfnamefont {J.~M.}\ \bibnamefont
  {Dudley}}\ and\ \bibinfo {author} {\bibfnamefont {J.~R. J.~R.}\ \bibnamefont
  {Taylor}},\ }\href@noop {} {\emph {\bibinfo {title} {{Supercontinuum
  generation in optical fibers}}}}\ (\bibinfo  {publisher} {Cambridge :
  Cambridge University Press},\ \bibinfo {address} {Cambridge},\ \bibinfo
  {year} {2010})\BibitemShut {NoStop}%
\bibitem [{\citenamefont {Gordon}(1986)}]{Gordon1986TheoryShift}%
  \BibitemOpen
  \bibfield  {author} {\bibinfo {author} {\bibfnamefont {J.~P.}\ \bibnamefont
  {Gordon}},\ }\href {\doibase 10.1364/ol.11.000662} {\bibfield  {journal}
  {\bibinfo  {journal} {Optics Letters}\ }\textbf {\bibinfo {volume} {11}},\
  \bibinfo {pages} {662} (\bibinfo {year} {1986})}\BibitemShut {NoStop}%
\bibitem [{\citenamefont {Kodama}\ and\ \citenamefont
  {Hasegawa}(1987)}]{Kodama1987NonlinearGuide}%
  \BibitemOpen
  \bibfield  {author} {\bibinfo {author} {\bibfnamefont {Y.}~\bibnamefont
  {Kodama}}\ and\ \bibinfo {author} {\bibfnamefont {A.}~\bibnamefont
  {Hasegawa}},\ }\href {\doibase 10.1109/JQE.1987.1073392} {\bibfield
  {journal} {\bibinfo  {journal} {IEEE Journal of Quantum Electronics}\
  }\textbf {\bibinfo {volume} {23}},\ \bibinfo {pages} {510} (\bibinfo {year}
  {1987})}\BibitemShut {NoStop}%
\bibitem [{\citenamefont {Skryabin}\ \emph {et~al.}(2003)\citenamefont
  {Skryabin}, \citenamefont {Luan}, \citenamefont {Knight},\ and\ \citenamefont
  {Russell}}]{Skryabin2003SolitonFibers}%
  \BibitemOpen
  \bibfield  {author} {\bibinfo {author} {\bibfnamefont {D.~V.}\ \bibnamefont
  {Skryabin}}, \bibinfo {author} {\bibfnamefont {F.}~\bibnamefont {Luan}},
  \bibinfo {author} {\bibfnamefont {J.~C.}\ \bibnamefont {Knight}}, \ and\
  \bibinfo {author} {\bibfnamefont {P.~S.~J.}\ \bibnamefont {Russell}},\ }\href
  {\doibase 10.1126/science.1088516} {\bibfield  {journal} {\bibinfo  {journal}
  {Science}\ }\textbf {\bibinfo {volume} {301}},\ \bibinfo {pages} {1705}
  (\bibinfo {year} {2003})}\BibitemShut {NoStop}%
\bibitem [{\citenamefont {Biancalana}\ \emph {et~al.}(2004)\citenamefont
  {Biancalana}, \citenamefont {Skryabin},\ and\ \citenamefont
  {Yulin}}]{Biancalana2004TheoryFibers}%
  \BibitemOpen
  \bibfield  {author} {\bibinfo {author} {\bibfnamefont {F.}~\bibnamefont
  {Biancalana}}, \bibinfo {author} {\bibfnamefont {D.~V.}\ \bibnamefont
  {Skryabin}}, \ and\ \bibinfo {author} {\bibfnamefont {A.~V.}\ \bibnamefont
  {Yulin}},\ }\href {\doibase 10.1103/PhysRevE.70.016615} {\bibfield  {journal}
  {\bibinfo  {journal} {Physical Review E}\ }\textbf {\bibinfo {volume} {70}},\
  \bibinfo {pages} {016615} (\bibinfo {year} {2004})}\BibitemShut {NoStop}%
\bibitem [{\citenamefont {Skryabin}\ and\ \citenamefont
  {Gorbach}(2010)}]{Skryabin2010TheoryWaves}%
  \BibitemOpen
  \bibfield  {author} {\bibinfo {author} {\bibfnamefont {D.~V.}\ \bibnamefont
  {Skryabin}}\ and\ \bibinfo {author} {\bibfnamefont {A.~V.}\ \bibnamefont
  {Gorbach}},\ }in\ \href {\doibase DOI: 10.1017/CBO9780511750465.010} {\emph
  {\bibinfo {booktitle} {Supercontinuum Generation in Optical Fibers}}},\
  \bibinfo {editor} {edited by\ \bibinfo {editor} {\bibfnamefont {J.~M.}\
  \bibnamefont {Dudley}}\ and\ \bibinfo {editor} {\bibfnamefont {J.~R.}\
  \bibnamefont {Taylor}}}\ (\bibinfo  {publisher} {Cambridge University
  Press},\ \bibinfo {address} {Cambridge},\ \bibinfo {year} {2010})\ pp.\
  \bibinfo {pages} {178--198}\BibitemShut {NoStop}%
\bibitem [{\citenamefont {Gorbach}\ and\ \citenamefont
  {Skryabin}(2008)}]{Gorbach2008SolitonFibers}%
  \BibitemOpen
  \bibfield  {author} {\bibinfo {author} {\bibfnamefont {A.~V.}\ \bibnamefont
  {Gorbach}}\ and\ \bibinfo {author} {\bibfnamefont {D.~V.}\ \bibnamefont
  {Skryabin}},\ }\href {\doibase 10.1364/OE.16.004858} {\bibfield  {journal}
  {\bibinfo  {journal} {Optics Express}\ }\textbf {\bibinfo {volume} {16}},\
  \bibinfo {pages} {4858} (\bibinfo {year} {2008})}\BibitemShut {NoStop}%
\bibitem [{\citenamefont {Guo}\ \emph {et~al.}(2015)\citenamefont {Guo},
  \citenamefont {Zhou}, \citenamefont {Steinert}, \citenamefont {Setzpfandt},
  \citenamefont {Pertsch}, \citenamefont {Chung}, \citenamefont {Chen},\ and\
  \citenamefont {Bache}}]{Guo2015SupercontinuumMatching}%
  \BibitemOpen
  \bibfield  {author} {\bibinfo {author} {\bibfnamefont {H.}~\bibnamefont
  {Guo}}, \bibinfo {author} {\bibfnamefont {B.}~\bibnamefont {Zhou}}, \bibinfo
  {author} {\bibfnamefont {M.}~\bibnamefont {Steinert}}, \bibinfo {author}
  {\bibfnamefont {F.}~\bibnamefont {Setzpfandt}}, \bibinfo {author}
  {\bibfnamefont {T.}~\bibnamefont {Pertsch}}, \bibinfo {author} {\bibfnamefont
  {H.-p.}\ \bibnamefont {Chung}}, \bibinfo {author} {\bibfnamefont {Y.-H.}\
  \bibnamefont {Chen}}, \ and\ \bibinfo {author} {\bibfnamefont
  {M.}~\bibnamefont {Bache}},\ }\href {\doibase 10.1364/OL.40.000629}
  {\bibfield  {journal} {\bibinfo  {journal} {Optics Letters}\ }\textbf
  {\bibinfo {volume} {40}},\ \bibinfo {pages} {629} (\bibinfo {year}
  {2015})}\BibitemShut {NoStop}%
\bibitem [{\citenamefont {Phillips}\ \emph {et~al.}(2011)\citenamefont
  {Phillips}, \citenamefont {Langrock}, \citenamefont {Pelc}, \citenamefont
  {Fejer}, \citenamefont {Hartl},\ and\ \citenamefont
  {Fermann}}]{Phillips2011SupercontinuumWaveguides}%
  \BibitemOpen
  \bibfield  {author} {\bibinfo {author} {\bibfnamefont {C.~R.}\ \bibnamefont
  {Phillips}}, \bibinfo {author} {\bibfnamefont {C.}~\bibnamefont {Langrock}},
  \bibinfo {author} {\bibfnamefont {J.~S.}\ \bibnamefont {Pelc}}, \bibinfo
  {author} {\bibfnamefont {M.~M.}\ \bibnamefont {Fejer}}, \bibinfo {author}
  {\bibfnamefont {I.}~\bibnamefont {Hartl}}, \ and\ \bibinfo {author}
  {\bibfnamefont {M.~E.}\ \bibnamefont {Fermann}},\ }\href {\doibase
  10.1364/OE.19.018754} {\bibfield  {journal} {\bibinfo  {journal} {Opt.
  Express}\ }\textbf {\bibinfo {volume} {19}},\ \bibinfo {pages} {18754}
  (\bibinfo {year} {2011})}\BibitemShut {NoStop}%
\bibitem [{\citenamefont {Guo}\ \emph {et~al.}(2013)\citenamefont {Guo},
  \citenamefont {Zeng}, \citenamefont {Zhou},\ and\ \citenamefont
  {Bache}}]{Guo2013NonlinearMedia}%
  \BibitemOpen
  \bibfield  {author} {\bibinfo {author} {\bibfnamefont {H.}~\bibnamefont
  {Guo}}, \bibinfo {author} {\bibfnamefont {X.}~\bibnamefont {Zeng}}, \bibinfo
  {author} {\bibfnamefont {B.}~\bibnamefont {Zhou}}, \ and\ \bibinfo {author}
  {\bibfnamefont {M.}~\bibnamefont {Bache}},\ }\href {\doibase
  10.1364/JOSAB.30.000494} {\bibfield  {journal} {\bibinfo  {journal} {Journal
  of the Optical Society of America B}\ }\textbf {\bibinfo {volume} {30}},\
  \bibinfo {pages} {494} (\bibinfo {year} {2013})}\BibitemShut {NoStop}%
\bibitem [{\citenamefont {Poberaj}\ \emph {et~al.}(2012)\citenamefont
  {Poberaj}, \citenamefont {Hu}, \citenamefont {Sohler},\ and\ \citenamefont
  {G{\"{u}}nter}}]{Poberaj2012LithiumDevices}%
  \BibitemOpen
  \bibfield  {author} {\bibinfo {author} {\bibfnamefont {G.}~\bibnamefont
  {Poberaj}}, \bibinfo {author} {\bibfnamefont {H.}~\bibnamefont {Hu}},
  \bibinfo {author} {\bibfnamefont {W.}~\bibnamefont {Sohler}}, \ and\ \bibinfo
  {author} {\bibfnamefont {P.}~\bibnamefont {G{\"{u}}nter}},\ }\href {\doibase
  10.1002/lpor.201100035} {\bibfield  {journal} {\bibinfo  {journal} {Laser
  {\&} Photonics Reviews}\ }\textbf {\bibinfo {volume} {6}},\ \bibinfo {pages}
  {488} (\bibinfo {year} {2012})}\BibitemShut {NoStop}%
\bibitem [{\citenamefont {Boes}\ \emph {et~al.}(2018)\citenamefont {Boes},
  \citenamefont {Corcoran}, \citenamefont {Chang}, \citenamefont {Bowers},\
  and\ \citenamefont {Mitchell}}]{Boes2018StatusCircuits}%
  \BibitemOpen
  \bibfield  {author} {\bibinfo {author} {\bibfnamefont {A.}~\bibnamefont
  {Boes}}, \bibinfo {author} {\bibfnamefont {B.}~\bibnamefont {Corcoran}},
  \bibinfo {author} {\bibfnamefont {L.}~\bibnamefont {Chang}}, \bibinfo
  {author} {\bibfnamefont {J.}~\bibnamefont {Bowers}}, \ and\ \bibinfo {author}
  {\bibfnamefont {A.}~\bibnamefont {Mitchell}},\ }\href {\doibase
  10.1002/lpor.201700256} {\bibfield  {journal} {\bibinfo  {journal} {Laser
  {\&} Photonics Reviews}\ }\textbf {\bibinfo {volume} {12}},\ \bibinfo {pages}
  {1700256} (\bibinfo {year} {2018})}\BibitemShut {NoStop}%
\bibitem [{\citenamefont {Rowe}\ \emph {et~al.}(2019)\citenamefont {Rowe},
  \citenamefont {Skryabin},\ and\ \citenamefont
  {Gorbach}}]{Rowe2019TemporalNanowaveguides}%
  \BibitemOpen
  \bibfield  {author} {\bibinfo {author} {\bibfnamefont {W.~R.}\ \bibnamefont
  {Rowe}}, \bibinfo {author} {\bibfnamefont {D.~V.}\ \bibnamefont {Skryabin}},
  \ and\ \bibinfo {author} {\bibfnamefont {A.~V.}\ \bibnamefont {Gorbach}},\
  }\href {\doibase 10.1103/PhysRevResearch.1.033146} {\bibfield  {journal}
  {\bibinfo  {journal} {Physical Review Research}\ }\textbf {\bibinfo {volume}
  {1}},\ \bibinfo {pages} {033146} (\bibinfo {year} {2019})}\BibitemShut
  {NoStop}%
\bibitem [{\citenamefont {Buryak}\ \emph {et~al.}(2002)\citenamefont {Buryak},
  \citenamefont {Trapani}, \citenamefont {Skryabin},\ and\ \citenamefont
  {Trillo}}]{Buryak2002}%
  \BibitemOpen
  \bibfield  {author} {\bibinfo {author} {\bibfnamefont {A.~V.}\ \bibnamefont
  {Buryak}}, \bibinfo {author} {\bibfnamefont {P.~D.}\ \bibnamefont {Trapani}},
  \bibinfo {author} {\bibfnamefont {D.~V.}\ \bibnamefont {Skryabin}}, \ and\
  \bibinfo {author} {\bibfnamefont {S.}~\bibnamefont {Trillo}},\ }\href
  {\doibase 10.1016/S0370-1573(02)00196-5} {\bibfield  {journal} {\bibinfo
  {journal} {Physics Reports}\ }\textbf {\bibinfo {volume} {370}},\ \bibinfo
  {pages} {63} (\bibinfo {year} {2002})}\BibitemShut {NoStop}%
\bibitem [{\citenamefont {Schiek}(1993)}]{Schiek1993NonlinearStructures}%
  \BibitemOpen
  \bibfield  {author} {\bibinfo {author} {\bibfnamefont {R.}~\bibnamefont
  {Schiek}},\ }\href {\doibase 10.1364/JOSAB.10.001848} {\bibfield  {journal}
  {\bibinfo  {journal} {Journal of the Optical Society of America B}\ }\textbf
  {\bibinfo {volume} {10}},\ \bibinfo {pages} {1848} (\bibinfo {year}
  {1993})}\BibitemShut {NoStop}%
\bibitem [{\citenamefont {Buryak}\ \emph {et~al.}(1995)\citenamefont {Buryak},
  \citenamefont {Trillo},\ and\ \citenamefont
  {Kivshar}}]{Buryak1995OpticalNonlinearities}%
  \BibitemOpen
  \bibfield  {author} {\bibinfo {author} {\bibfnamefont {A.~V.}\ \bibnamefont
  {Buryak}}, \bibinfo {author} {\bibfnamefont {S.}~\bibnamefont {Trillo}}, \
  and\ \bibinfo {author} {\bibfnamefont {Y.~S.}\ \bibnamefont {Kivshar}},\
  }\href {\doibase 10.1364/OL.20.001961} {\bibfield  {journal} {\bibinfo
  {journal} {Optics Letters}\ }\textbf {\bibinfo {volume} {20}},\ \bibinfo
  {pages} {1961} (\bibinfo {year} {1995})}\BibitemShut {NoStop}%
\bibitem [{\citenamefont {Bache}\ and\ \citenamefont
  {Schiek}(2012)}]{Bache2012ReviewResponse}%
  \BibitemOpen
  \bibfield  {author} {\bibinfo {author} {\bibfnamefont {M.}~\bibnamefont
  {Bache}}\ and\ \bibinfo {author} {\bibfnamefont {R.}~\bibnamefont {Schiek}},\
  }\href {https://arxiv.org/pdf/1211.1721.pdf} {\bibfield  {journal} {\bibinfo
  {journal} {preprint arXiv:1211.1721}\ } (\bibinfo {year} {2012})}\BibitemShut
  {NoStop}%
\bibitem [{\citenamefont {Cai}\ \emph {et~al.}(2018)\citenamefont {Cai},
  \citenamefont {Gorbach}, \citenamefont {Wang}, \citenamefont {Hu},\ and\
  \citenamefont {Ding}}]{Cai2018HighlyNano-waveguides}%
  \BibitemOpen
  \bibfield  {author} {\bibinfo {author} {\bibfnamefont {L.}~\bibnamefont
  {Cai}}, \bibinfo {author} {\bibfnamefont {A.~V.}\ \bibnamefont {Gorbach}},
  \bibinfo {author} {\bibfnamefont {Y.}~\bibnamefont {Wang}}, \bibinfo {author}
  {\bibfnamefont {H.}~\bibnamefont {Hu}}, \ and\ \bibinfo {author}
  {\bibfnamefont {W.}~\bibnamefont {Ding}},\ }\href {\doibase
  10.1038/s41598-018-31017-0} {\bibfield  {journal} {\bibinfo  {journal}
  {Scientific Reports}\ }\textbf {\bibinfo {volume} {8}},\ \bibinfo {pages}
  {12478} (\bibinfo {year} {2018})}\BibitemShut {NoStop}%
\bibitem [{\citenamefont {Gorbach}\ and\ \citenamefont
  {Ivanov}(2016)}]{Gorbach2016PerturbationGeneration}%
  \BibitemOpen
  \bibfield  {author} {\bibinfo {author} {\bibfnamefont {A.~V.}\ \bibnamefont
  {Gorbach}}\ and\ \bibinfo {author} {\bibfnamefont {E.}~\bibnamefont
  {Ivanov}},\ }\href {\doibase 10.1103/PhysRevA.94.013811} {\bibfield
  {journal} {\bibinfo  {journal} {Physical Review A}\ }\textbf {\bibinfo
  {volume} {94}},\ \bibinfo {pages} {013811} (\bibinfo {year}
  {2016})}\BibitemShut {NoStop}%
\bibitem [{\citenamefont {Gorbach}\ and\ \citenamefont
  {Skryabin}(2007)}]{Gorbach2007TheoryFibers}%
  \BibitemOpen
  \bibfield  {author} {\bibinfo {author} {\bibfnamefont {A.~V.}\ \bibnamefont
  {Gorbach}}\ and\ \bibinfo {author} {\bibfnamefont {D.~V.}\ \bibnamefont
  {Skryabin}},\ }\href {\doibase 10.1103/PhysRevA.76.053803} {\bibfield
  {journal} {\bibinfo  {journal} {Physical Review A - Atomic, Molecular, and
  Optical Physics}\ }\textbf {\bibinfo {volume} {76}},\ \bibinfo {pages}
  {053803} (\bibinfo {year} {2007})}\BibitemShut {NoStop}%
\bibitem [{\citenamefont {Kivshar}\ \emph
  {et~al.}(2003{\natexlab{b}})\citenamefont {Kivshar}, \citenamefont
  {Agrawal},\ and\ \citenamefont {Agrawal}}]{Kivshar2003ChapterSolitons}%
  \BibitemOpen
  \bibfield  {author} {\bibinfo {author} {\bibfnamefont {Y.~S.}\ \bibnamefont
  {Kivshar}}, \bibinfo {author} {\bibfnamefont {G.~P.}\ \bibnamefont
  {Agrawal}}, \ and\ \bibinfo {author} {\bibfnamefont {G.~P.}\ \bibnamefont
  {Agrawal}},\ }in\ \href {\doibase
  https://doi.org/10.1016/B978-012410590-4/50002-4} {\emph {\bibinfo
  {booktitle} {Optical Solitons}}},\ \bibinfo {editor} {edited by\ \bibinfo
  {editor} {\bibfnamefont {Y.~S.}\ \bibnamefont {Kivshar}}, \bibinfo {editor}
  {\bibfnamefont {G.~P.}\ \bibnamefont {Agrawal}}, \ and\ \bibinfo {editor}
  {\bibfnamefont {G.~P.}\ \bibnamefont {Agrawal}}}\ (\bibinfo  {publisher}
  {Academic Press},\ \bibinfo {address} {Burlington},\ \bibinfo {year} {2003})\
  pp.\ \bibinfo {pages} {31--62}\BibitemShut {NoStop}%
\bibitem [{\citenamefont {Skryabin}\ and\ \citenamefont
  {Yulin}(2005)}]{Skryabin2005TheoryFibers}%
  \BibitemOpen
  \bibfield  {author} {\bibinfo {author} {\bibfnamefont {D.~V.}\ \bibnamefont
  {Skryabin}}\ and\ \bibinfo {author} {\bibfnamefont {A.~V.}\ \bibnamefont
  {Yulin}},\ }\href {\doibase 10.1103/PhysRevE.72.016619} {\bibfield  {journal}
  {\bibinfo  {journal} {Physical Review E}\ }\textbf {\bibinfo {volume} {72}},\
  \bibinfo {pages} {016619} (\bibinfo {year} {2005})}\BibitemShut {NoStop}%
\bibitem [{\citenamefont {Rowe}\ \emph {et~al.}()\citenamefont {Rowe},
  \citenamefont {Skryabin},\ and\ \citenamefont
  {Gorbach}}]{RoweDatasetHttps://doi.org/10.15125/BATH-00808}%
  \BibitemOpen
  \bibfield  {author} {\bibinfo {author} {\bibfnamefont {W.}~\bibnamefont
  {Rowe}}, \bibinfo {author} {\bibfnamefont {D.}~\bibnamefont {Skryabin}}, \
  and\ \bibinfo {author} {\bibfnamefont {A.}~\bibnamefont {Gorbach}},\ }\href
  {\doibase 10.15125/BATH-00808} {\enquote {\bibinfo {title} {{Dataset for
  "Raman solitons in waveguides with simultaneous quadratic and Kerr
  nonlinearities". Bath: University of Bath Research Data Archive.
  https://doi.org/10.15125/BATH-00808}},}\ }\BibitemShut {NoStop}%
\end{thebibliography}%

\end{document}